\documentclass[12pt,aps,prb]{revtex4}
\usepackage{graphicx}
\begin{document}
\pagestyle{plain}

\vspace*{0.7cm}
\begin{center}
{\bf Microwave Heating of Water, Ice and Saline Solution:
Molecular Dynamics Study}  

\vspace*{1.3cm}
Motohiko Tanaka, and Motoyasu Sato \\
Coordinated Research Center, National Institute for Fusion 
Science, Toki 509-5292, Japan 

\vspace*{1.5cm}
{\bf Abstract} 
\end{center}

In order to study the heating process of water by the microwaves 
of 2.5-20GHz frequencies, we have performed molecular dynamics 
simulations by adopting a non-polarized water model that have 
fixed point charges on rigid-body molecules. 
All runs are started from the equilibrated states derived from 
the I$_{c}$ ice with given density and temperature. 
In the presence of microwaves, the molecules of liquid water exhibit 
rotational motion whose average phase is delayed from the microwave 
electric field.  
Microwave energy is transferred to the kinetic and inter-molecular 
energies of water, where one third of the absorbed microwave energy 
is stored as the latter energy. 
The water in ice phase is scarcely heated by microwaves because of 
the tight hydrogen-bonded network of water molecules. 
Addition of small amount of salt to pure water substantially 
increases the heating rate because of the weakening by defects 
in the water network due to sloshing large-size negative ions.

\vspace*{0.5cm}

\newpage
\section{Introduction}
\label{Sec.1}

It is well known that microwaves (300MHz - 300GHz range) can heat 
solid, liquid and gaseous matters such as engineering materials, 
laboratory plasmas, and biological matters including living cells. 
A microwave oven used in daily food processing is one of such
applications. 
However, unlike laser lights whose photon energy is from several 
to tens of electronvolts, the photon energy of microwaves is
as small as $ h \nu \sim 10^{-5} $eV and their period is
by orders of magnitude longer than the electronic processes 
(a few fs) occurring in molecules.
Nevertheless, microwaves can control chemical reactions, 
synthesize organic and inorganic materials 
\cite{kappe,appl-syn}, and sinter metal oxides with high 
energy efficiency \cite{Roy}.
Thus, for the microwave-related heating and reactions to take 
place, non-resonant interactions between waves and materials 
that persist for many wave periods are expected.

The physical and chemical properties of matters are the basis 
of material science, engineering and biological applications. 
Especially, the dielectric permittivity was measured 
extensively as the function of electromagnetic wave 
frequency and temperature \cite{data-hip}, 
and precise experimental formulas were presented for the 
dielectric properties of water, including enhanced heating 
of salty water \cite{water-diel}. 
The imaginary part of the dielectric permittivity is related 
to the energy absorption by dielectric medium, which is 
termed as {\it dielectric loss}. 
For water, the rotational relaxation of permanent electric 
dipoles occurs approximately in 8ps at 25$^{\circ}$C 
(frequency of 120GHz) \cite{dip-relax}.
For the microwaves less than and around this frequency, only 
translational and rotational motions of polar molecules are 
expected to respond to the waves.

The diffusion coefficients and dielectric relaxation properties
of water, i.e. the response of electric dipoles to a given 
initial impulse, were studied theoretically \cite{hirata}.
Numerically, the melting of ice at normal pressure was 
investigated with several water models \cite{melting}.
The heating and diffusion of water under high-frequency
microwaves and infra-red electromagnetic waves were investigated 
by molecular dynamics simulations using elaborated point-charge 
models that incorporated either charge polarization or molecular
flexibility for the system of 256 water molecules \cite{Eng1,Eng2}. 
Precise studies showed that the polarizable water model TIP4P-FQ
combined with the Lekner method gives more accurate results than
fixed-charge model, including the potential energy, dipole moment, 
dielectric constant and relaxation times \cite{Eng3}. 
However, the polarizable model requires large computation times, 
roughly three times more than fixed-charge models with the Ewald 
method for 256 water molecules. It becomes progressively more 
demanding as O($ N^{2}$) with the increasing number of molecules 
compared to O($ N^{3/2}$) of the Ewald method.
Since we treat large systems with 2700 water molecules with or
without salt ions, we adopt the non-polarized water model SPC/E 
combined with the Ewald method, which yields rather qualitative 
but reasonable results for electrostatic properties. 

Complementary to the previous studies, we examine the heating 
of water by low-frequency microwaves in typical phases, 
including liquid water, ice and dilute saline solution, by means of 
classical molecular dynamics simulations \cite{MD-DNA}. 
We use the microwaves in the 2.5-20GHz frequency range and 
the field strength of $ E_{rms} \sim $0.01V/\AA, and 
a relatively large system containing 2700 water molecules. 
As we focus on the heating process in the low frequency and low
field strength regime, we adopt a fixed point-charge, 
rigid-body model for water and a constant volume periodic 
system without an energy or particle reservoir. 
As for the diagnosis, we directly measure the kinetic and 
potential energies to obtain transferred energies 
from the microwaves. 

We have confirmed that water in the liquid phase is heated by 
microwaves through the excitation of rotational motion 
of permanent electric dipoles of water molecules which is delayed
from the wave electric field.
The microwave energy is transferred to kinetic and internal 
(inter-molecular) energies of water, where the energy stored
as the internal one amounts to about one third that of the 
total absorbed power.
Dilute salt water is heated significantly more rapidly than 
pure water because of the weakening of hydrogen-bonded water 
network due to large-size salt ions including Cl$^{-}$.
On the other hand, the water in ice phase is hardly heated by
low-frequency microwaves, since electric dipoles exhibit
substantial directional inertia due to the rigid network of 
water molecules.

This paper is organized as follows. 
In the next section, the simulation method and parameters adopted 
in this study are described. In Sec.3, our simulation results of 
microwave heating are presented for three typical cases including
ice, liquid water and salt-added water.  
Sec.4 will be a summary of this paper.

\vspace*{-0.3cm}
\section{Simulation Method and Parameters}
\label{Sec.2}

To study the heating process of water, we use the microwaves 
of the frequency range 2.5-20GHz, whose wavelengths and 
periods are 1.5-12cm and 50-400ps, respectively.  
These scale-lengths are much larger than our model water system 
whose edge length is approximately 40\AA \ in one direction 
(typically one molecule resides in every 3\AA). 
Also, all the involved velocities are much less than 
the speed of light ($ v/c \ll 1 $). Under such circumstances, we 
can safely assume with the motion of water molecules and salt
ions that the microwave is represented by spatially 
uniform, time alternating electric field of the form 
\begin{eqnarray}
\label{eq:micro-E}
 && \tilde{E}_{x}(t) = E_{0} \sin \omega t , 
    \ \ \tilde{B} = 0. 
\end{eqnarray}

For the purpose of our study mentioned above, the water in liquid 
and ice phases is approximated by the assembly of 
three-point-charge rigid-body molecules known as the SPC/E model 
\cite{SPC}.
These molecules are placed in a constant volume cubic box. 
In this model, fractional charges $ \delta q $ are distributed 
at the hydrogen sites and $ -2 \delta q $ at the oxygen site,
where $ \delta q= 0.424|e| $ with $ e $ the electronic charge.
The $i$-th atom of the water molecule moves under the Coulombic 
and Lennard-Jones forces
\begin{eqnarray}
\label{eq:motion}
 && m_{i} \frac{d{\bf v}_{i}}{dt} 
   = -\nabla \left\{ 
   \sum_{j} \frac{q_{i}q_{j}}{r_{ij}}
   + 4 \epsilon_{LJ}
           \left[ \left(\frac{\sigma}{r_{ij}} \right)^{12} -
                  \left(\frac{\sigma}{r_{ij}} \right)^{6} \right]
   \right\}, \\
 && \frac{d{\bf x}_{i}}{dt} = {\bf v}_{i},
\end{eqnarray}
where $ {\bf r}_{i} $ and $ {\bf v}_{i} $ are the position and 
velocity of the $ i $-th atom, respectively, 
$ r_{ij}= |{\bf r}_{i} - {\bf r}_{j}| $, 
$ m_{i} $ and $ q_{i} $ are its mass and charge,
respectively.  For the Lennard-Jones potential between interacting 
$ i $-th and $ j $-th atoms, the combination rules
\begin{eqnarray}
 && \sigma= (\sigma_{i} + \sigma_{j})/2, \ 
    \epsilon_{LJ}= \sqrt{\epsilon_{i} \epsilon_{i}}
\end{eqnarray}
are used where $ \sigma_{i} $ and $ \epsilon_{i} $ are the 
diameter and Lennard-Jones potential depth of the $ i $-th atom,
respectively.  In Eq.(\ref{eq:motion}), the Lennard-Jones force 
is calculated only for the oxygen atoms with 
$ \sigma_{O}= $ 3.17\AA \ and $ \epsilon_{LJ,O}=$ 0.65kJ/mol.
The length of the hydrogen-oxygen bond is 1.00\AA \ and the
angle between these bonds is 109.47 degrees.  
The constraint dynamics procedure called "Shake and Rattle algorithm"
\cite{shake} is used to maintain these bond length and angle. 
Salt Na$^{+}$Cl$^{-} $ is added in some runs. The Lennard-Jones
forces are calculated for the pair of the ionic atom and 
oxygen atom of water molecules, where the diameter and the 
Lennard-Jones potential depth used for Na$^{+}$ are 2.57\AA \ 
and 0.062kJ/mol, respectively, and those for Cl$^{-} $ are 
4.45\AA \ and 0.446kJ/mol, respectively. 

As for the initial conditions, we use highly structured crystal ice 
I$_{c} $ with 14 molecules in all three directions under given 
density 1.00g/cm$^{3}$ for liquid water (the edge length of 
the cubic box is fixed at $ L= $43.448\AA), and 0.93g/cm$^{3}$ 
for ice ($ L= $44.512\AA).  
Thus, we have approximately 2700 water molecules in the box
with periodic boundary conditions.
Water molecules, especially those of ice and liquid water, 
form a three-dimensional network with hollow unit cells
consisting of six-membered water rings \cite{ohmine}. 
The connection between the molecules is made by the hydrogen 
bond between a hydrogen atom and an adjacent oxygen atom 
that belongs to a different molecule. 
Each molecule donates two hydrogen bonds to adjacent molecules 
and accepts two hydrogen bonds from other molecules to form 
the hydrogen-bonded network, thus there is a freedom to which 
neighboring molecules two hydrogen atoms are donated from
one molecule. 
The initial orientations of water molecules are prescribed
such that the directions of two hydrogen atoms bonded to an oxygen 
atom are swapped under the I$_{c} $ symmetry until the sum of the 
bonded O-H vectors is nullified for each of all three directions; 
the number of water molecules in any direction must be an even number
\cite{iceIc}, which is 14 for the system length 44.512\AA. 
All simulation runs are preceded by an equilibration phase of 
100ps at given temperature and density before microwaves are applied.  

The calculation of the Coulombic forces under the periodic boundary
condition requires the charge sum in the first Brillouin zone
and their infinite number of mirror images (the Ewald sum\cite{Ewald}).
The sum is calculated with the use of the particle-particle,
particle-mesh algorithm and the procedure of minimizing the rms error 
in the force \cite{Eastwood,Deserno}. 
We use $ (32)^{3} $ spatial meshes and
the 3rd order spline interpolation for the calculation of the 
reciprocal space contributions to the Coulombic forces, 
and take direct summations for the short-range particle-particle 
forces with the real-space cutoff 10\AA \ and the Ewald screening 
parameter $ \alpha= 0.25 $\AA$^{-1}$. 
This yields the maximum rms error in the force $ 1.1 \times 10^{-4} $. 
We note that the accuracy of the Ewald sum can be optimized with
the Lekner method for highly ionic environments at the expense of
computation times \cite{Eng5}.
The time integration step is $ \Delta t= $ 1fs, for which
the detection level of heating for liquid water is 
$ (dW/dt)_{noise} \sim 2 \times 10^{-5} kT_{0}$/ps.
We use our PC cluster machines, each of which consists of four 
Pentium 4/EM64T (3.4GHz) processors, and a 500ps run takes 
typically 48 hours.

Three types of runs are performed, namely, 
(i) ice of the I$_{c}$ form at 230K, 
(ii) pure water whose initial temperature is 300K, 
and (iii) salt water with 1mol\% NaCl (3.2wt\% or 0.57M) 
starting at 300K. 
In numerical simulations, we use large electric fields 
in order to detect heating in a reasonable computation time, 
which is typically $ E_{0} \sim 1 \times 10^{6} $V/cm.  
Nevertheless, this is not a large electric field when viewed from 
water molecules, since the dipole energy $ E_{0}p_{0} 
\sim 4.8 \times 10^{-3} $eV ($ p_{0} \sim 2.4 \times 10^{-18} $esu 
cm \ is the electric dipole of our model water) is still less 
than thermal energy $ kT_{0} \sim 2.6 \times 10^{-2} $eV 
at room temperature $ T_{0}= 300 $K, where $ k $ is the 
Boltzmann constant. The dipole energy is by orders of magnitude 
less than the hydrogen-bond energy 2.5eV per bond.

\vspace*{-0.2cm}
\section{Simulation Results}
\label{Sec.3}


In the following sections, unless otherwise specified, we use
the microwaves of the frequency 10GHz (period 100ps) and 
the strength $ E_{0}= 2.23 \times 10^{6} $V/cm, which
corresponds to the ratio of the dipole energy and thermal 
energy at $ T_{0}= 300 $K, $ E_{0}p_{0}/kT_{0} \sim 0.42 $.

\subsection{Heating of Crystal Ice}
\label{Sec.3a}

The microwave heating of water in ice phase is examined 
for the crystal ice I$_{c} $ at the temperature 230K.
Figure 1 shows the time history of average quantities:
(a) the kinetic energy of water molecules $ W_{kin}=
< \hspace*{-0.1cm} \frac{1}{2}m {\bf v}^{2} 
   \hspace*{-0.1cm}> $, which includes both translational 
and rotational energies, (b) the Lennard-Jones energy, 
$ W_{LJ}= \ <\hspace*{-0.1cm} 4 \epsilon_{LJ} 
   [(\sigma/r_{ij})^{12} - (\sigma/r_{ij})^{6}] 
          \hspace*{-0.1cm}> $,
and (c) the inter-molecular energy, which is the 
sum of the Lennard-Jones energy and the Coulombic energy
$ W_{c}= \ (1/N(N-3)) \sum_{ij} ' q_{i}q_{j}/r_{ij} $, 
where the prime means the summation excluding the charge 
pairs on the same molecule. 
The kinetic energy above is the index of heating, and
the maximum and minimum of Fig.1(a) correspond to
230$ \pm $7K, respectively.
Despite the microwave application for $ t > 0 $, the kinetic 
and inter-molecular energies remain nearly constant except for 
small periodic fluctuations.

The geometrical arrangement of water molecules is shown in 
Fig.2 at the final time $ t $=500ps for (a) ice at 230K, 
(b) liquid water initially at 300K (to be mentioned in Sec.IIIB),
and (c)-(e) their enlarged plots of the edge parts.
The water in ice phase of Fig.2(c) has a complete crystal 
structure, which remains nearly intact after the microwave 
application as seen in Fig.2(d) and (a).
The distribution functions of electric dipoles of water 
molecules for the ice in terms of the directional cosine 
\begin{eqnarray}
\label{eq:cos-theta}
 && \cos \Theta_{i} = \hat{x} \cdot {\bf p}_{i}/ \it{p}_{i},
\end{eqnarray}
are shown in Fig.3, which also stay almost unchanged during 
the microwave application (the ordinate is in a logarithmic scale), 
where $ {\bf p}_{i}(t) $ is the electric dipole of the $ i $-th 
molecule, $ \hat{x} $ is the unit vector in the $ x $ direction. 
This immobility of electric dipoles is attributed to the rigidity 
of the network of water molecules due to hydrogen bonds. 
These observations indicate that pure crystal ice is not heated by 
microwaves of a few GHz frequency at the applied intensity.
On the other hand, methane hydrates were found to be raptured 
by microwave fields at higher intensities \cite{Eng4}.

\vspace*{-0.2cm}
\subsection{Heating of Liquid Water}
\label{Sec.3b}

Now we examine the microwave heating of water in the liquid 
phase starting at the temperature 300K.
In the cubic box there are 2744 water molecules which have been 
equilibrated without the microwave field at this temperature.

The depiction of molecular arrangement of Fig.2(b) and (e) 
at $ t= 500$ps after microwave application clearly shows that 
the water molecules in the liquid phase are randomized both in guiding 
center positions and molecular orientations by absorbing microwaves. 
Figure 4 shows the time history of (a) the kinetic energy
of water molecules, (b) the Lennard-Jones energy, and 
(c) the sum of the Coulombic and Lennard-Jones energies. 
When microwaves are switched on at t=0 in Fig.4, the kinetic 
energy of water molecules begins to increase at a constant 
rate; the final temperature is 350K.
The average distance between the molecules is expanding 
during this rearrangement of molecules (the energy minimum of
the Lennard-Jones potential is located at $ r_{OO}= 
2^{1/6}\sigma_{O}= $3.56\AA).
We note that the water at rest is in a minimum energy state 
because of strong attraction forces due to hydrogen bonds.
The decrease in the Lennard-Jones energy and the increase in the 
Coulombic energy take place simultaneously under the microwave
field, but the latter is larger than the former, thus the 
inter-molecular energy increases.  
The microwave energy is transferred to kinetic and 
inter-molecular energies of water, where the latter is about 
35\% that of the total absorbed power, as seen by 
comparing Fig.4(a) and (c). 

The observed heating is attributed to the excitation of the 
rotational motion of permanent electric dipoles of water
molecules and simultaneous energy transfer to translational 
energy via molecular collisions. 
Angular distributions of electric dipoles as the function of
the directional cosine of their orientation are shown 
in Fig.5 in 50ps intervals. 
Before the microwave application, the electric dipoles 
in Fig.5(a) have random orientations at room temperature. 
When the microwave electric field is present, the dipoles 
align along the direction of the time-alternating electric 
field $ \tilde{E} $, as seen in Fig.5(b) and (c). 
At each instant, the angular distribution of electric dipoles 
nearly follows the statistical (Boltzmann) distribution, 
\begin{eqnarray}
\label{eq:F(theta)}
 && F(\Theta) \cong A^{-1} \exp( \tilde{E} p \cos \Theta/kT) 
\end{eqnarray}
with regard to the angle $ \Theta $, where
$ \varepsilon_{d}= -\tilde{E} p \cos \Theta $ is the dipole 
energy and the normalization constant is
$ A= \int F(\Theta) sin\Theta d\Theta
= (kT/Ep)[\exp(kT/Ep)-\exp(-kT/Ep)] $.

The radial distribution function (RDF) between oxygen atoms 
$ g_{OO}(r) $ for liquid water in Fig.6(a) reveals that the
initial gap located at $ r_{OO} \sim $3.5\AA \ is filled 
after the microwave application. The gap at $ r_{OH} \sim 
$2.5\AA \ in the RDF between hydrogen and oxygen atoms 
$ g_{OH}(r) $ is also partially filled.  
These observations indicate the randomization of the 
orientation of electric dipoles and the increase in 
the distances between oxygen atoms. They are consistent 
with the increase in the Coulombic energy
$ \sum_{i>j} q_{i}q_{j}/r_{ij} $, where the major
contribution comes form the hydrogen and oxygen pairs 
due to hydrogen bonds. 
 
However, it should be noted that a finite phase difference 
is required between the orientation of electric dipoles of 
water molecules and the electric field for the energy 
to be transferred from microwaves to water. 
The sum of the $ x $-component of electric dipoles 
is obtained by
\begin{eqnarray}
\label{eq:Px}
 && P_{x}= \int_{-\pi}^{\pi} p \cos\Theta F(\Theta) 
    \sin\Theta d\Theta \\ 
 && \hspace*{0.6cm} \cong \frac{p}{2} \int_{-1}^{1} \cos\Theta 
   \left[ 1+\frac{Ep}{kT}\cos\Theta \right]
                          d\cos\Theta \nonumber \\ 
 &&  \hspace*{0.6cm} = \left( \frac{Ep}{3kT} 
                          \right) p \nonumber,
\end{eqnarray}
where we have used Eq.(\ref{eq:F(theta)}) and expanded it 
assuming $ Ep/kT \ll 1 $ (the next order term is 
$ (p/15)(Ep/kT)^{3} $, which is a few percent of the 
above leading term even for $ Ep/kT \sim 0.4 $). 
Then, we put the electric dipole
$ p(t)= p_{0} \sin(\omega(t-\tau)) $ with the phase lag $ \tau $. 
The work done to the dipoles at the position $ x $ 
per unit time by the electric field Eq.(\ref{eq:micro-E}) is,
\begin{eqnarray}
\label{eq:dW_wat/dt}
 && \frac{dW_{E}}{dt} = \left< E \frac{dP_{x}}{dt} \right> \\
 &&  \hspace*{0.6cm} = \frac{(p_{0}E_{0})^{2}}{3kT}
          <\hspace*{-0.1cm} \sin\omega t 
             \frac{d}{dt} \left\{ 
            \sin\omega t \sin^{2}(\omega(t-\tau)) \right\}
              \hspace*{-0.1cm}> \nonumber \\
 && \hspace*{0.6cm} = \frac{(p_{0}E_{0})^{2}}{24kT} 
            \omega \sin 2\omega\tau \nonumber
\end{eqnarray}
If we assume the Debye-type relaxation  
$ \tau \cong \zeta /2kT $ and $ \omega \tau \ll 1 $, where 
$ \zeta $ is the friction from adjacent water molecules 
\cite{water-relax}, then we have 
\begin{eqnarray}
\label{eq:dT/dt-an}
 && \frac{dW_{E}}{dt} \cong \frac{1}{24} \zeta \omega^{2} 
     \left( \frac{p_{0}E_{0}}{kT} \right)^{2}.
\end{eqnarray}
This formula indicates that, without friction $ \zeta= 0 $, 
the phase lag vanishes $ \tau = 0 $, thus we have no microwave 
heating of water.

The heating and energy absorption rates of liquid water at 300K
by the microwaves of f= 10GHz and $ E_{0}= 2.23 \times 10^{6} $V/cm  
are $ 7.6 \times 10^{-4} kT_{0}$/ps and 
$ 1.1 \times 10^{-3} kT_{0}$/ps, respectively (Table I). 
The expected energy absorption rate using the experimentally 
obtained imaginary part of the dielectric constant 
$ \epsilon"/\epsilon_{0} \cong 13 $ is 
$ \omega \epsilon" E^{2}/8\pi n_{0} \sim 
5.4 \times 10^{-5} $erg/ps/molecule $ = 
1.3 \times 10^{-3} kT_{0} $/ps/molecule, where $ n_{0} \cong 
3.3 \times 10^{22} $molecules/cm$^{3} $. 
This estimate is in fair agreement with the simulation value
of the energy absorption rate.
Further, by equating Eq.(\ref{eq:dW_wat/dt}) and the energy 
absorption rate in the simulation, then we have the estimate
for the phase lag as $ \tau \sim 19 $ps.

The total energy absorption rate by the water system may be 
obtained by integrating the local energy absorption rate over 
the depth ($ x $-coordinate). Since the wave electric field 
attenuates as $ E(x)= E(0) \exp(-x/\lambda) $ in dielectric 
medium, then we have
\begin{eqnarray}
 && \int_{0}^{\ell} dx = \lambda \int_{E_{min}}^{E(0)} dE/E,
\end{eqnarray}
where $ E_{min}= E(0) \exp(-\ell/\lambda) $, and $ E(0) $ is 
the electric field at the interface (inside) of water. 
The length $ \lambda $ is a few cm for 2-10GHz microwave 
in water.
For the power dependence on the electric field $ \dot{W}_{E}
\sim E^{\alpha} $, the integration yields the same power 
dependence even when $ \ell \ge \lambda $ by substituting 
$ E_{0} $ of the local formula by 
$ (\lambda / \ell(\alpha-1)) E(0) $.

Existence of the phase-lag is verified in Fig.7 which shows 
the temporal phase variations of the $ x $-component of the
average water electric dipole $ P_{x}(t) $ and the electric 
field, for (a) ice at 230K and (b) liquid water initially at 300K.  
For the ice case, variations of the electric dipoles are
very small.
For the liquid water, on the other hand, we see large
oscillations in the $ x $-component of electric dipoles.
Their amplitude is almost the value expected by
Eq.(\ref{eq:Px}), and a finite phase difference exists 
between the electric dipole and the electric field. 
The phase-lag of electric dipoles from the microwave 
electric field is about 12ps on average.  
This is about two third that of the phase lag obtained 
above using Eq.(\ref{eq:dW_wat/dt}), but is
close to the rotational relaxation time 8ps of water 
dipoles at 25$^{\circ} $C, and a fraction of the macroscopic 
relaxation time 40ps (25GHz) of bulk water \cite{wat-relax}. 

The dependence of the energy transfer rate to liquid water 
on the strength of the microwave electric field is shown in 
Fig.8, for the temperature 300K and the wave frequency 10GHz. 
Here, the time rate of the increase in the kinetic energy is 
plotted by filled circles, and that in the total energy
(the sum of the kinetic and inter-molecular energies) is plotted
by open circles, with both axes in logarithmic scales.
These two energy transfer rates from the microwaves increase 
by power laws of the microwave electric field, and are 
scaled as 
\begin{eqnarray}
\label{eq:dT/dt-E-sim}
 && \left( \frac{dW}{dt} \right)_{wat} \propto E^{\alpha} , 
\end{eqnarray}
with $ \alpha \cong 2.0 $ both for the wave power absorbed by 
the water system and the kinetic energy of water. 
We note that about 30\% of the absorbed microwave energy is stored
as the inter-molecular energy to rearrange water molecules.
These dependences on the electric field agree with that 
given by Eq.(\ref{eq:dT/dt-an}). 

The dependence of the energy transfer rate on the frequency 
of microwaves is shown in Fig.9, where the frequency is 
in the $ f= \omega/2\pi $= 2.5-20GHz range. 
The transferred energy from the microwaves increases with 
the wave frequency, and are scaled by
\begin{eqnarray}
\label{eq:dT/dt-f-sim}
 &&  \left( \frac{dW}{dt} \right)_{wat} 
        \propto \omega^{\beta},
\end{eqnarray}
with $ \beta \cong 1.5 $ both for the total energy 
absorbed by the water system and for the kinetic energy 
of water molecules. 
These frequency dependences are somewhat weaker than 
that of the formula Eq.(\ref{eq:dT/dt-an}), as the 
approximation $ \sin 2\omega \tau \cong 2\omega \tau $ 
may deteriorate at high frequencies. 

In Table I, the heating rates that we have measured by numerical
simulations of the ice and liquid water at two initial temperatures 
are listed, together with that for the salt water
(to be mentioned in Sec.IIIC). 
The heating rate for the ice is below our detection level.
The heating rate of hot water at 400K (in liquid phase, 
because of the constant volume) is small and only a fraction 
in comparison with that at room temperature. 
This is due to less friction at higher temperatures, since
the inter-molecular distances of water molecules become large. 
The simulation value is consistent with the experiment 
\cite{data-hip}.

\vspace*{-0.2cm}
\subsection{Heating of Salt Water}
\label{Sec.3c}

We examine the heating process of dilute saline solution, which 
corresponds to our daily applications including the heating 
of salty food in a microwave oven. 
In the simulation, we place 27 Na$^{+}$ and 27 Cl$^{-}$ ions 
(1mol\% NaCl or 3.2wt\% salinity) at random positions to the water 
system at room temperature.
The number of water molecules here 2717 is less than that of the 
pure water case with the same volume to avoid overlap of salt ions 
and water molecules.
We equilibrate the solution for 100ps before the microwave 
application at $ t=0 $.  

Figure 10 shows the time history of salt water heating by microwaves
of 10GHz and the field strength $ 2.23 \times 10^{6} $V/cm 
($ E_{0}p_{0}/kT_{\rm 300} \sim 0.42 $).
The kinetic energy of water molecules increases only slightly 
faster (11\%) than for pure water in the early stage.
After a waiting time of 0.8ns, a rapid heating phase sets in and 
the salt water is heated by a few times better than the pure water. 
The length of this microscopic waiting time among the runs with
different microwave field strength is not indexed to 
the water temperature at which the rapid heating sets in. 
The microwave power is absorbed also as the inter-molecular 
energy, which is comparable to the increase in the kinetic energy 
(Fig.10(c)).

The behavior of salt ions in saline solution after the microwave 
application is shown in Fig.11. Both the cations and anions 
are accelerated by the electric field, and the envelops (maximum
values) of their average velocities expand continuously in time.  
Their acceleration is nearly in-phase and out-of-phase with the 
microwave field for cations and anions, respectively.
Interestingly, the degree of the acceleration is appreciably 
larger for the heavier and larger anions than cations.  
This fact may be due to that small-size cations are well contained 
in the water network compared with anions whose diameter is
comparable to the cell size of the network (this point is
to be mentioned in the RDF of salt ions in Fig.12).
The mean displacements of salt ions in Fig.11 shows that 
anions shift toward the negative $ x $-direction and its
amplitude is larger than that of cations which shift to
the positive $ x $-direction.
We note that the amount of the positional shift for anions 
increases linearly up to $ t= $0.8ns. At the end of this
period when the rapid hating phase begins, the positional shift 
is roughly 7\AA \ which is comparable to the unit cell size of 
the water network consisting of hydrogen-bonded six-membered 
rings. This reveals that the rapid heating of saline solution 
is connected with sloshing of anions among the neighboring cells. 

The velocity distribution functions along the direction of
the microwave electric field has been examined for salt ions
in saline solution. 
Unlike the solids such as zeolites and titanium oxides
\cite{Eng5,zeolite}, the velocity distribution function
consists of a single Boltzmann distribution and supra-thermal 
component has not been detected under the wave 
frequency and intensity of the present study. 

The radial distribution functions (RDF) of oxygen-oxygen pairs 
$ g_{OO}(r) $ and those of oxygen-hydrogen pairs $ g_{OH}(r) $ 
are shown for the above salt water in Fig.6(b). 
Initial RDFs are shown by solid lines and those at the final 
times are shown by shaded histograms. 
We see drastic differences between the RDFs before and after
the microwave application.
The height of the first peak at 1.9\AA \ in $ g_{OH}(r) $ 
and that at 2.8\AA \ in $ g_{OO}(r) $ are much reduced, and
the gaps between the first and second peaks are filled.
This reveals the randomization of the position and orientation of
water molecules, and thus weakening of the water network.

The accumulated RDF for the oxygen-hydrogen pairs at the initial 
time has a flat pedestal, from which the association number of 
a hydrogen atom to oxygen atoms is approximately two at $ r 
\cong $ 2\AA. 
At $ t= $1.4ns, the inner edge of the accumulated RDF
is receded for oxygen-hydrogen pairs.  
The accumulated RDF of the Na$^{+}$ ions $ G_{Na^{+}O}(r) $ in 
Fig.12 shows that they are associated with five oxygen atoms 
at $ r \sim $2.5\AA \ which changes only slightly during
the microwave application. 
This means that Na$^{+}$ ions are well contained in the unit
cell of the water network, as previously mentioned.
On the other hand, $ G_{Cl^{-}O}(r) $ shows that a Cl$^{-} $ ion 
is loosely associated with seven oxygen atoms at $ r \cong $ 3.8\AA. 
After the microwave application at $ t= $1.4 ns, the inner edges 
of the accumulated RDFs for Cl$^{-}$ ions, $ G_{Cl^{-}O}(r) $ 
and $ G_{Cl^{-}H}(r) $ recede both with respect to
oxygen and hydrogen atoms.
This implies that large Cl$^{-} $ ions are constantly pressed by
water molecules and are less stably trapped in the water network. 
As has been suggested by Fig.11, oscillating Cl$^{-} $ ions 
are generating friction against water molecules and deteriorating 
the hydrogen bonding between them.

The heating rate of the saline solution with 1mol\% NaCl 
in the present molecular dynamics simulations and that of 
experiments are also listed in Table I.
The experimental values are very diverse probably because of
the different experimental devices and sample sizes. 
The old experiments were made with a Pt foil-wrapped 
resonator in which samples were placed \cite{data-hip}. 
Another measurement was done with a cavity resonator using 
two microwaves of different frequencies, one for heating and 
the other for measurement \cite{data-fuk}.
The heating rate of salt water in the simulation ($ t > 1$ns) 
is just between the experimental values.

The dependence of the energy transfer rate from microwaves 
to salt water on the frequency of microwaves is shown in Fig.13.
Both the energy transfer rate to the system (open triangles) 
and that to the kinetic energy of water (filled triangles)
are scaled as 
\begin{eqnarray}
\label{dT/dt-omg-salt}
 && \left( \frac{dW}{dt} \right)_{salt} 
                         \propto \omega^{\gamma} ,
\end{eqnarray}
with $ \gamma \cong 0.6 $, which is less sensitive to the
microwave frequency than for pure water.
%
%
%
We expect that the heating of salt water is not
attributed to a simple acceleration of salt ions but is the 
result of the interactions between charged salt ions and 
the water network.

In order to verify the aforementioned statement, we prepare
special saline solutions with (i) heavy salt ions, (ii) small 
Cl ions $ \sigma_{Cl^{-}}= \sigma_{Na^{+}} $= 2.57\AA, and 
(iii) {\it non-charged} salt, then apply the microwaves of 10GHz 
frequency and the strength $ E_{0}= 2.23 \times 10^{6} $V/cm. 
In the case (i), the salt ions are made ten times as heavy as their
original masses, thus their vibrations are slower and smaller 
than the normal salt. 
The heating rate of water is found to be $ dW_{kin}/dt \sim 
1.5 \times 10^{-3} kT_{0}$/ps, which is similar to that of the 
normal salt water shown in Table I.
For the case (iii), the heating rate is small,
$ dW_{kin}/dt \sim 7.1 \times 10^{-4} kT_{0}$/ps, and this is
almost the same as that of the pure water.  
This reveals that a non-charged sphere, even if it is as large
as the unit cell of the water network, does not actively interact 
with water molecules. 
An interesting case is the small salt ions of (ii), for which
the heating rate is intermediate between those of the pure 
and salt waters, $ dW_{kin}/dt \sim 9.9 \times 10^{-4} kT_{0}$/ps.
These special runs clearly indicate that large-size, charged ions 
like Cl$^{-} $ are playing important roles in the water heating 
process.
Namely, large salt ions that do not fit in the unit cell of 
the water network make defects to it, and cleave the network 
through oscillations in response to the microwave electric field.

\vspace*{-0.3cm} 
\section{Summary}
\label{Sec.4}

In this paper, we have studied the heating process of liquid 
water, ice and dilute saline solution by microwaves of 2.5-20GHz, 
where molecular dynamics simulations are performed with 
the explicit point-charge, rigid-body water model.  
We have verified that water in the liquid phase is heated 
via the rotational excitation of water electric dipoles, 
which is delayed from the microwave electric field. 
Microwave energy is transferred to both kinetic and 
inter-molecular energies of water, where the former corresponds 
to the heating of water and the latter is stored internally 
to rearrange water molecules. 
The latter energy occupies about one third of the total
absorbed microwave power. 

Hot water is significantly less heated than the water at 
room temperature, because the electric dipoles follow
the microwave field with less phase-lags due to less friction.
Water in ice phase is scarcely heated because the electric 
dipoles cannot rotate due to a tightly hydrogen-bonded 
network of water molecules.
Dilute saline solution is substantially more rapidly heated
than pure water. 
This is due to cleavage and weakening of hydrogen-bonded 
water network by large-size Cl$^{-} $ salt ions oscillating
in the microwave electric field. 

We remark that the microwave electric field in usual 
applications may be several orders of magnitude smaller than 
that used in the present study. 
The actual microwave field within the experimental sample 
should be considerably small compared with the surrounding
vacuum space, because microwaves attenuate in the 
dielectric medium.
If we assume that the strength of the microwave electric 
field is 10V/cm (700W microwaves correspond to the electric 
field 730V/cm, which becomes $ 1/\epsilon(3GHz) \cong
1/77 $ in water), and the wave frequency is 2.5GHz, 
then using the scalings Eqs.(\ref{eq:dT/dt-E-sim}) 
and (\ref{eq:dT/dt-f-sim}) we have 
\begin{eqnarray}
\label{eq:dTkin/dt-real}
 && \frac{dW_{kin}}{dt} \cong 1.8 \times 10^{-3}
      kT_{0}/{\rm s}
\end{eqnarray}
for the heating rate of water, where $ T_{0}= 300 $K.  
This corresponds to the temperature increase of roughly 
70 degrees in two minutes, which is consistent with 
our daily experiences.

We have two other remarks: (i) For the microwave of small 
intensity, several orders of magnitude more wave periods 
are required to heat water and related materials. 
(ii) A part of the energy stored internally may be 
released as heat in a long time scale, but it is beyond
the scope our present study.
Finally, the energy associated with the electric dipole 
$ ({\bf E} \cdot {\bf p}) $ in the presence of microwaves 
in our simulation is not exceeding thermal energy or 
hydrogen-bond energy of water molecules. 
Therefore, except for secondary phenomena such as a long-time 
relaxation of water structures, our simulations are correctly 
reproducing the essence of the microwave heating process 
of water and saline solution. 

\begin{acknowledgments}
One of the authors (M.T.) thanks Prof.I.Ohmine and 
Dr.M.Matsumoto for fruitful discussions and providing him 
with the I$_{c} $ ice generating program. 
This work was supported by Grant-in-Aid No.16032217 (2003-2005) 
and No.18070005 (2006-2010) from the Ministry of Education, Science 
and Culture of Japan. The present simulations were performed using 
our Linux-based PC cluster machines comprising of Pentium 4/EM64T 
and Opteron 275 processors.
\end{acknowledgments}

\newpage

\newpage
\begin{table} 
\caption{The heating and energy absorption rates of liquid water, 
ice and saline solution at various initial temperatures, with 
experimentally measured dielectric losses as references. 
The microwave frequency in numerical simulations is 10GHz, the 
electric field strength is $ E_{0}= 2.23 \times 10^{6} $V/cm (or 
$ E_{0}p_{0}/kT_{0} \sim 0.42 $), and the heating and absorption 
rates are in the unit of kT$_{0}$/ps/molecule, with 
$ T_{0}= 300 $K.
a) Hot water in the liquid phase (due to a constant volume),
b) 1mol\% NaCl salt concentration.  
The dielectric losses by experiments are 
$ \epsilon''/ \epsilon_{0} $, for 2.45GHz and 25$^{\circ}$C, 
c) 95$^{\circ}$C \cite{data-hip}, 
d) 1.0mol\% salinity (6GHz) \cite{data-fuk}, 
and e) 0.5mol\% salinity (3GHz) \cite{data-hip}.
}
 
\vspace*{0.7cm}
\begin{tabular}{cclll} \hline \hline
\hspace{0.3cm} state \hspace{0.3cm} & temperature \hspace{0.3cm} & 
heating rate \hspace{0.3cm} & absorption rate \hspace{0.3cm} &
experiments \hspace{0.3cm} \\
\hline
 ice         & 230K & very small            & very small & $<$0.01 \\ 
 water       & 300K & $7.6 \times 10^{-4} $ & $1.1 \times 10^{-3}$ & 13 \\ 
 water$^{a}$ & 400K & $2.6 \times 10^{-4} $ & $4.7 \times 10^{-4}$ & 
2.4$^{c}$ \\ 
 \hspace*{0.5cm} salt water$^{b}$ & 300K & $1.7 \times 10^{-3} $ 
             & $3.0 \times 10^{-3}$ & 18$^{d}$, 42$^{e}$ \\ 
\hline \hline
\end{tabular}
\end{table}
\vspace*{7.7cm}

\newpage
\begin{figure}
\vspace*{-3.0cm}
\centerline{\scalebox{0.80}{\includegraphics{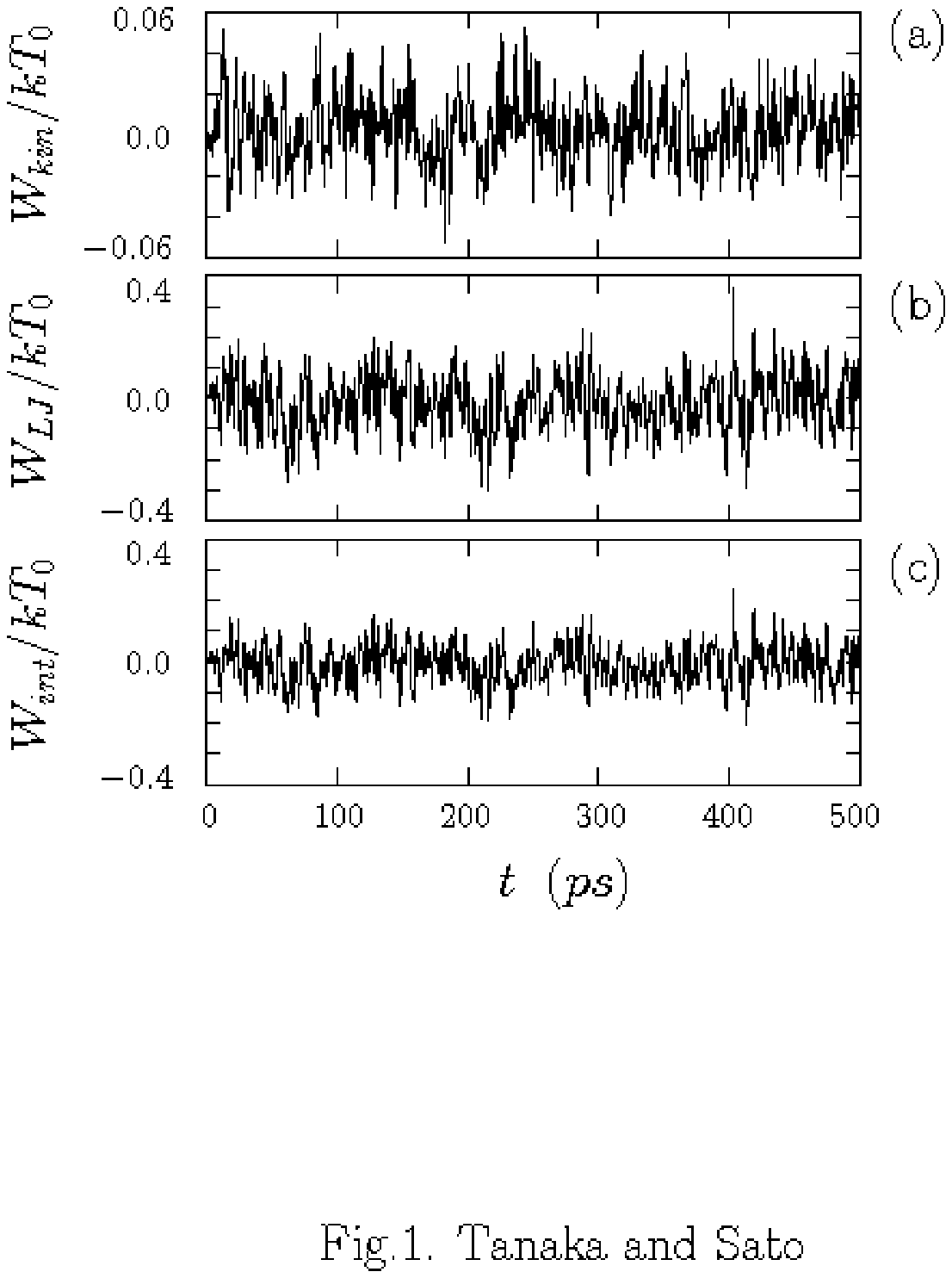}}}
\vspace*{-7.0cm}
\caption{
\noindent
The time history of (a) the average kinetic energy, (b) 
the average Lennard-Jones energy, and (c) the average 
inter-molecular energy per molecule, for the ice at temperature 230K. 
Microwaves are applied for $ t > 0 $, whose frequency is 10GHz 
and its strength is $ 2.23 \times 10^{6} $V/cm, or 
$ E_{0}p_{0}/kT_{\rm 300} \sim 0.42 $.
}
\label{Fig.1}
\end{figure}

\newpage
\begin{figure}
\centerline{\scalebox{1.1}{\includegraphics{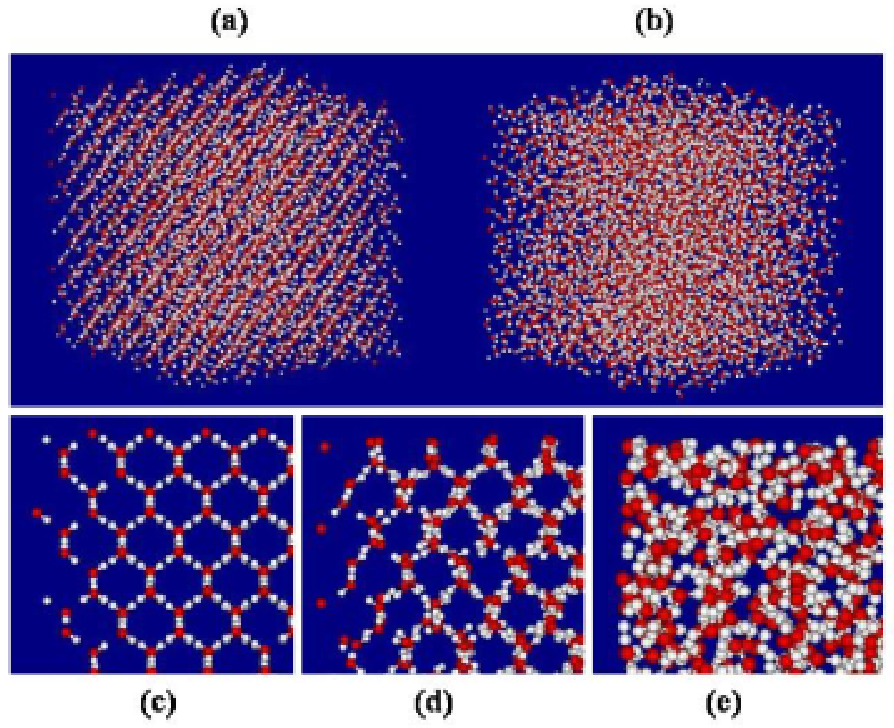}}}
\vspace*{1.0cm}
\caption{
\noindent
(color) The geometrical arrangement of water molecules 
at $ t $=500ps after the microwave application, for 
(a) the ice at 230K, and (b) liquid water initially at 300K. 
Enlarged edge parts are shown for (c) the initial I$_{c}$ ice,
(d) the ice at 230K (edge part of (a)), and (e) liquid water 
(edge part of (b)).
The microwave frequency is 10GHz and its field strength 
$ E_{0} $ is $ E_{0}p_{0}/kT_{0} \sim 0.42 $ with 
$ T_{0}= 300 $K.
}
\label{Fig.2}
\end{figure}

\newpage
\begin{figure}
\vspace*{-3.0cm}
\centerline{\scalebox{0.80}{\includegraphics{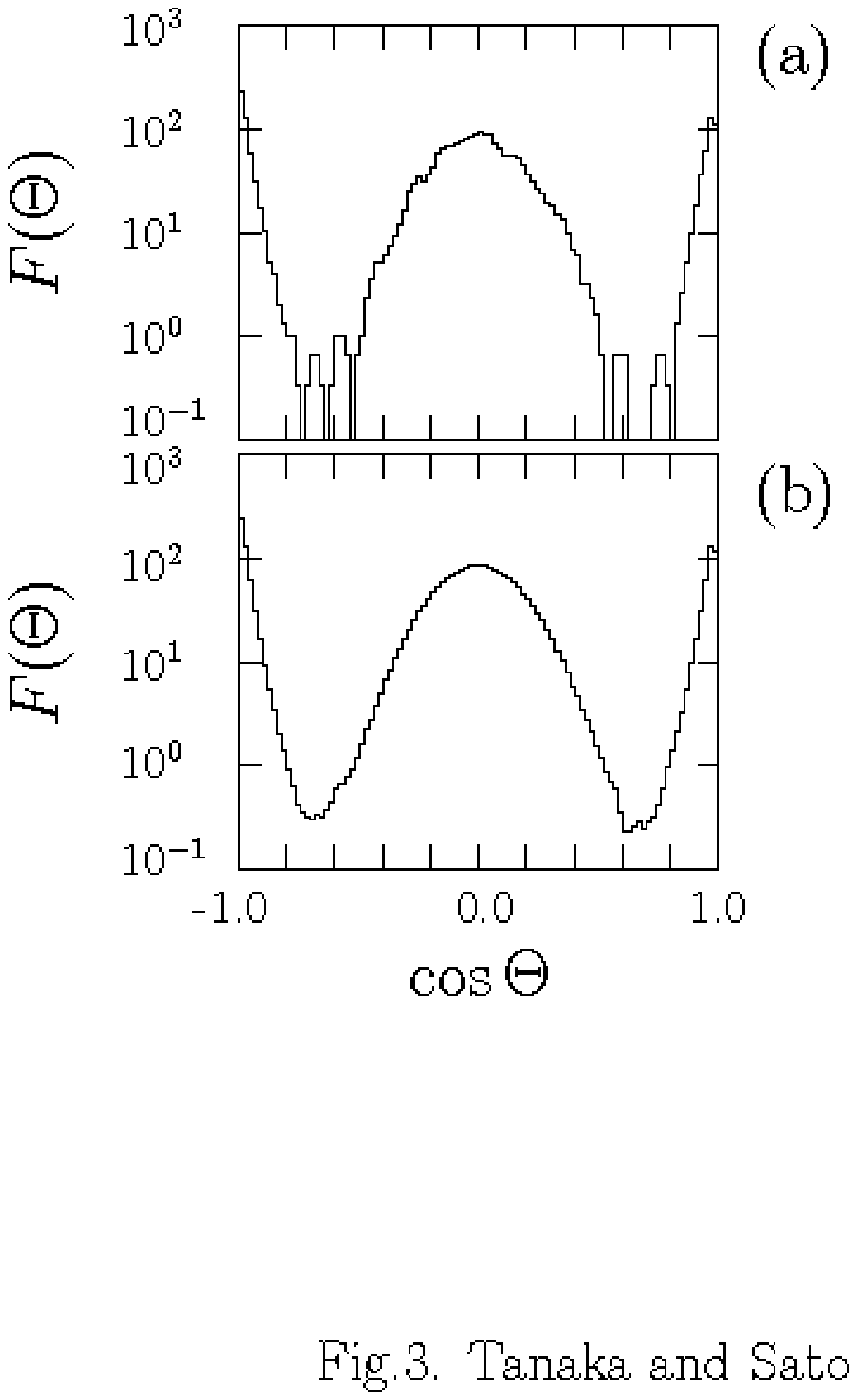}}}
\vspace*{-7.0cm}
\caption{
\noindent
The distribution of water electric dipoles as
the function of the directional cosine $ \cos \Theta $, 
Eq.(\ref{eq:cos-theta}), for the ice at temperature 230K
shown in Fig.1, (a) before (t=0) and (b) after (t=500ps) 
the application of microwaves.  
The ordinates are in a logarithmic scale.
}
\label{Fig.3}
\end{figure}

\newpage
\begin{figure}
\vspace*{-3.0cm}
\centerline{\scalebox{0.80}{\includegraphics{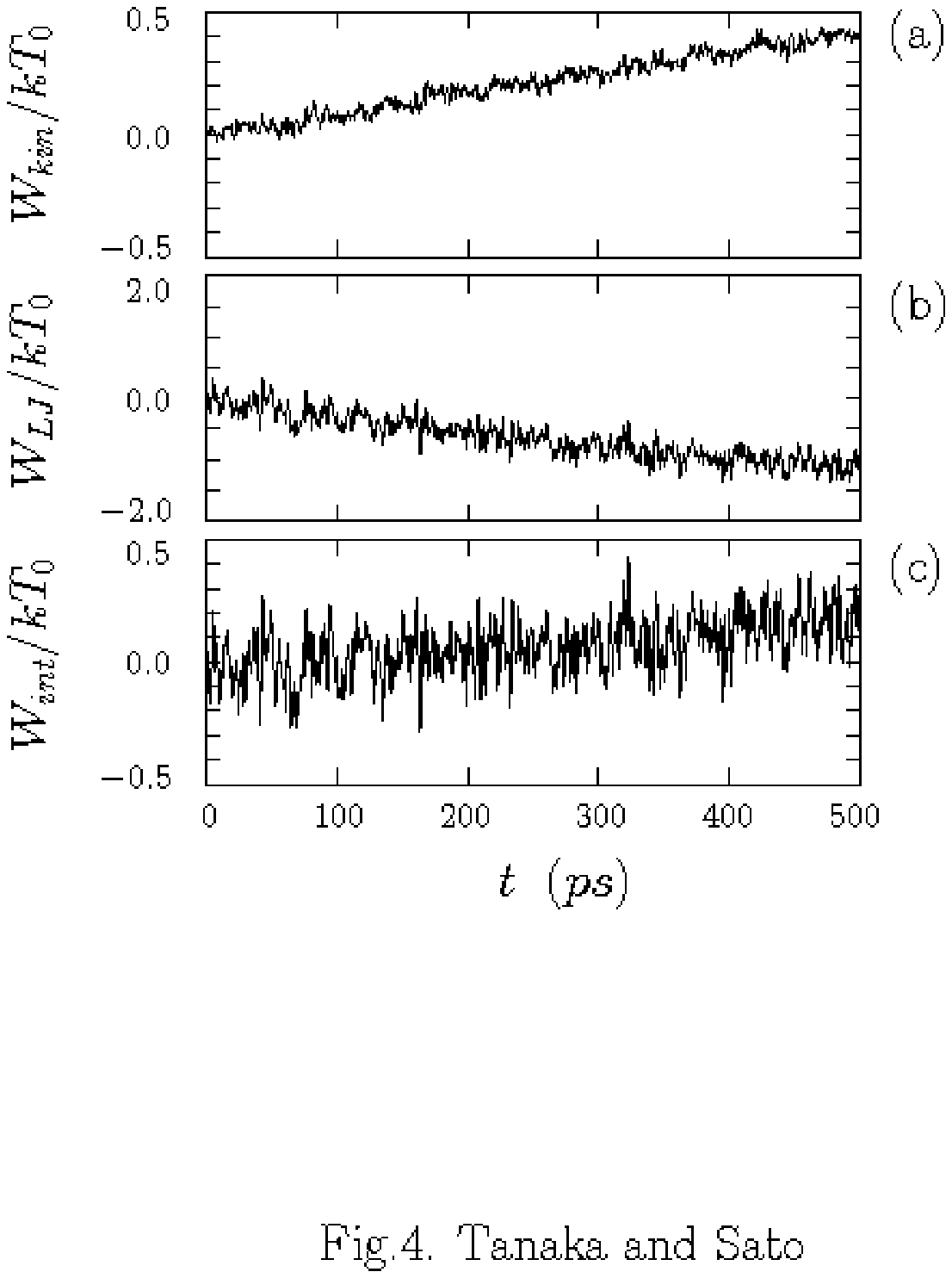}}}
\vspace*{-7.0cm}
\caption{
\noindent
The time history of (a) the average kinetic energy, (b) 
the average Lennard-Jones energy, and (c) the sum of average 
Coulombic and Lennard-Jones energies per molecule, for liquid water of 
initial temperature 300K.  
Microwave frequency is 10GHz and its strength is 
$ E_{0}p_{0}/kT_{0} \sim 0.42 $.
}
\label{Fig.4}
\end{figure}

\newpage
\begin{figure}
\vspace*{-3.0cm}
\centerline{\scalebox{0.80}{\includegraphics{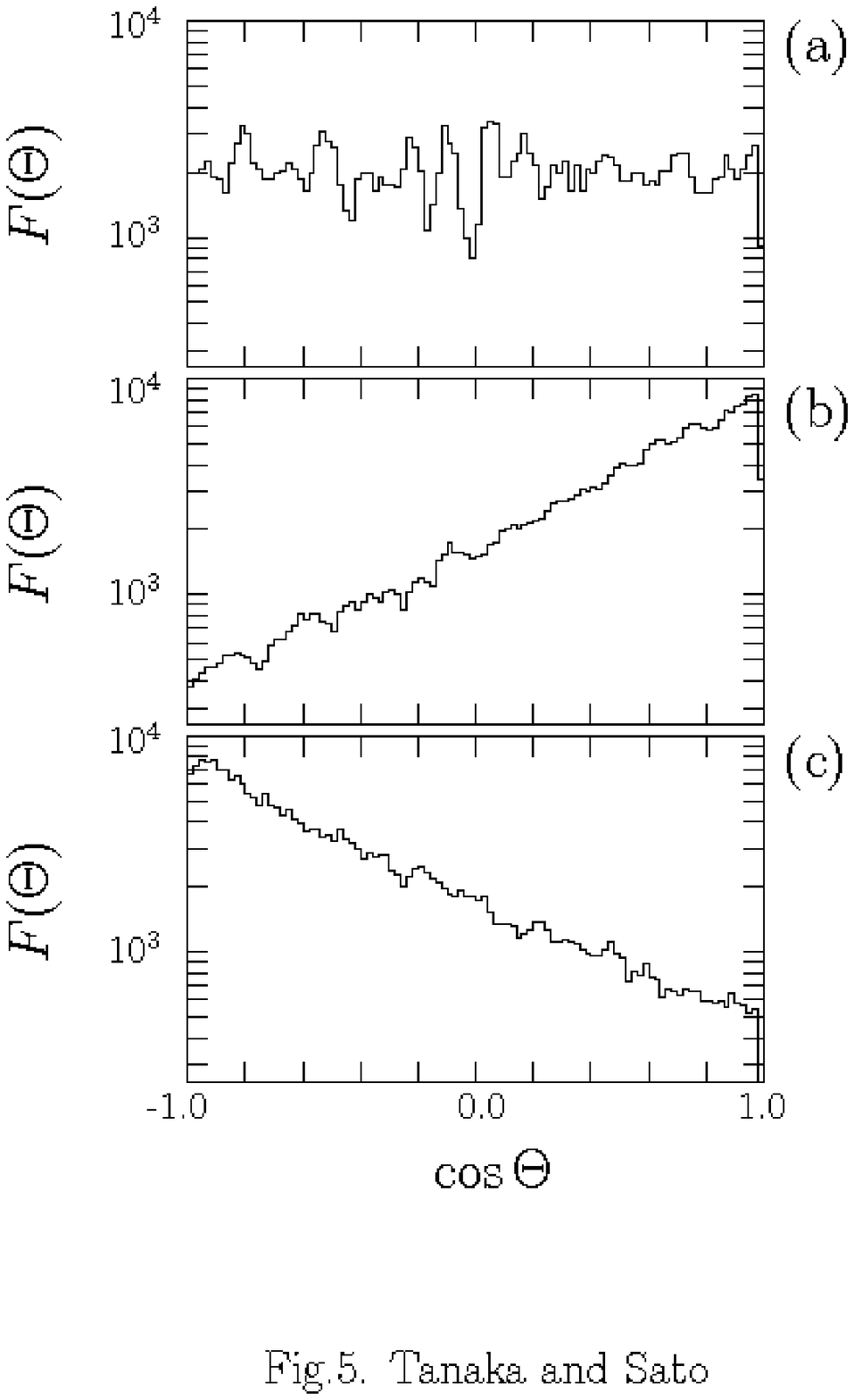}}}
\vspace*{-5.0cm}
\caption{
\noindent
The distribution of electric dipoles as the function
of the directional cosine $ \cos \Theta $, for (a) just before 
the application of microwaves (t=0), at (b) t=50ps, and 
(c) t=100ps for liquid water initially at temperature 300K 
shown in Fig.4.  The ordinates are in a logarithmic scale.
}
\label{Fig.5}
\end{figure}

\newpage
\begin{figure}
\vspace*{-2.0cm}
\centerline{\scalebox{0.80}{\includegraphics{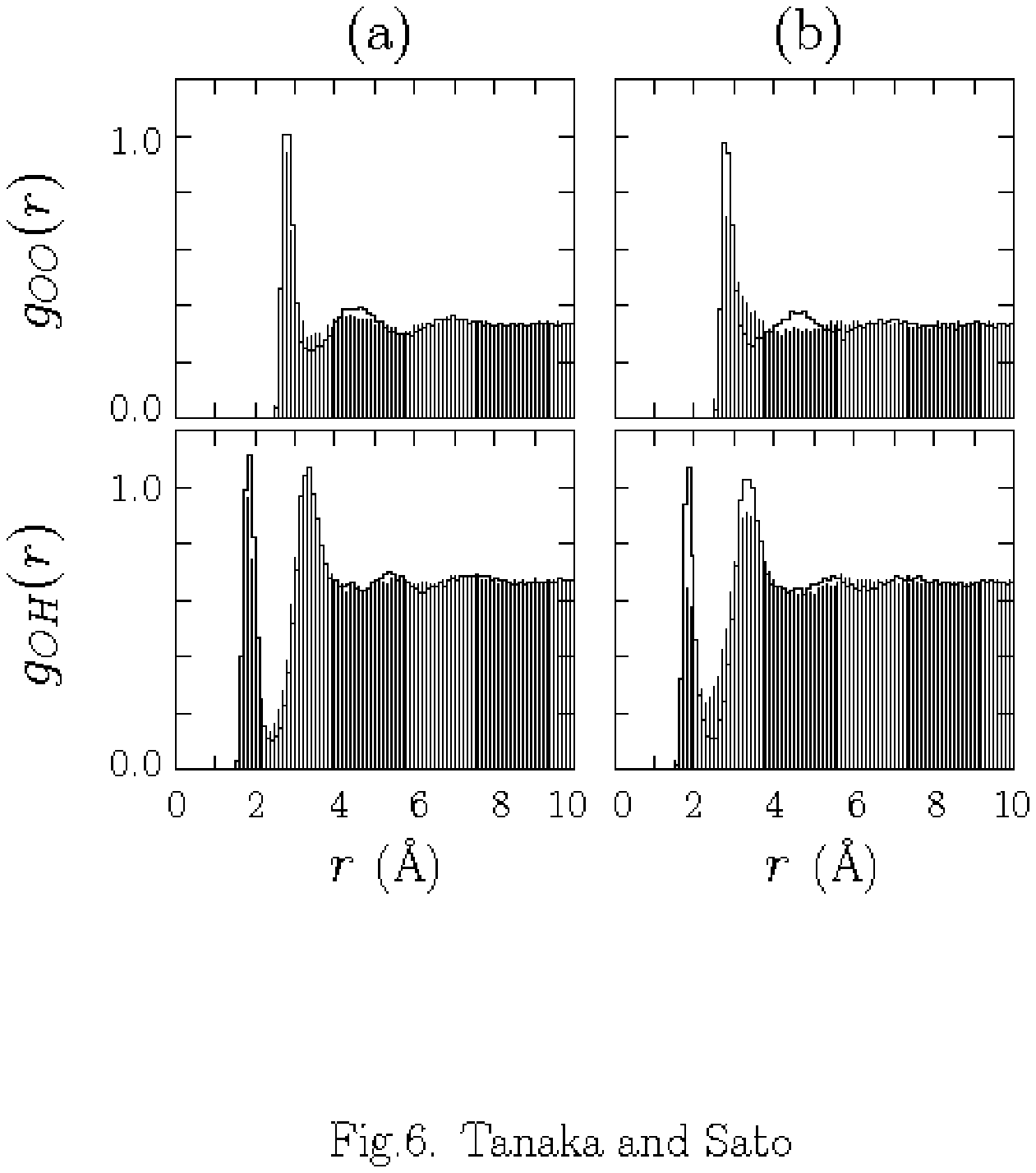}}}
\vspace*{-7.0cm}
\caption{
\noindent
The radial distribution functions (RDF) of the oxygen-oxygen 
pairs $ g_{OO}(r) $ and those of oxygen-hydrogen pairs 
$ g_{OH}(r) $, for (a) liquid water at 300K and (b) salt water 
with 1mol\% salinity.  The initial RDFs are shown by solid 
lines and those at the final times ($ t= $ 500ps for (a), 
and $ t= $ 1.4ns for (b)) are shown by shaded histograms.
}
\label{Fig.6}
\end{figure}

\newpage
\begin{figure}
\vspace*{-3.0cm}
\centerline{\scalebox{0.80}{\includegraphics{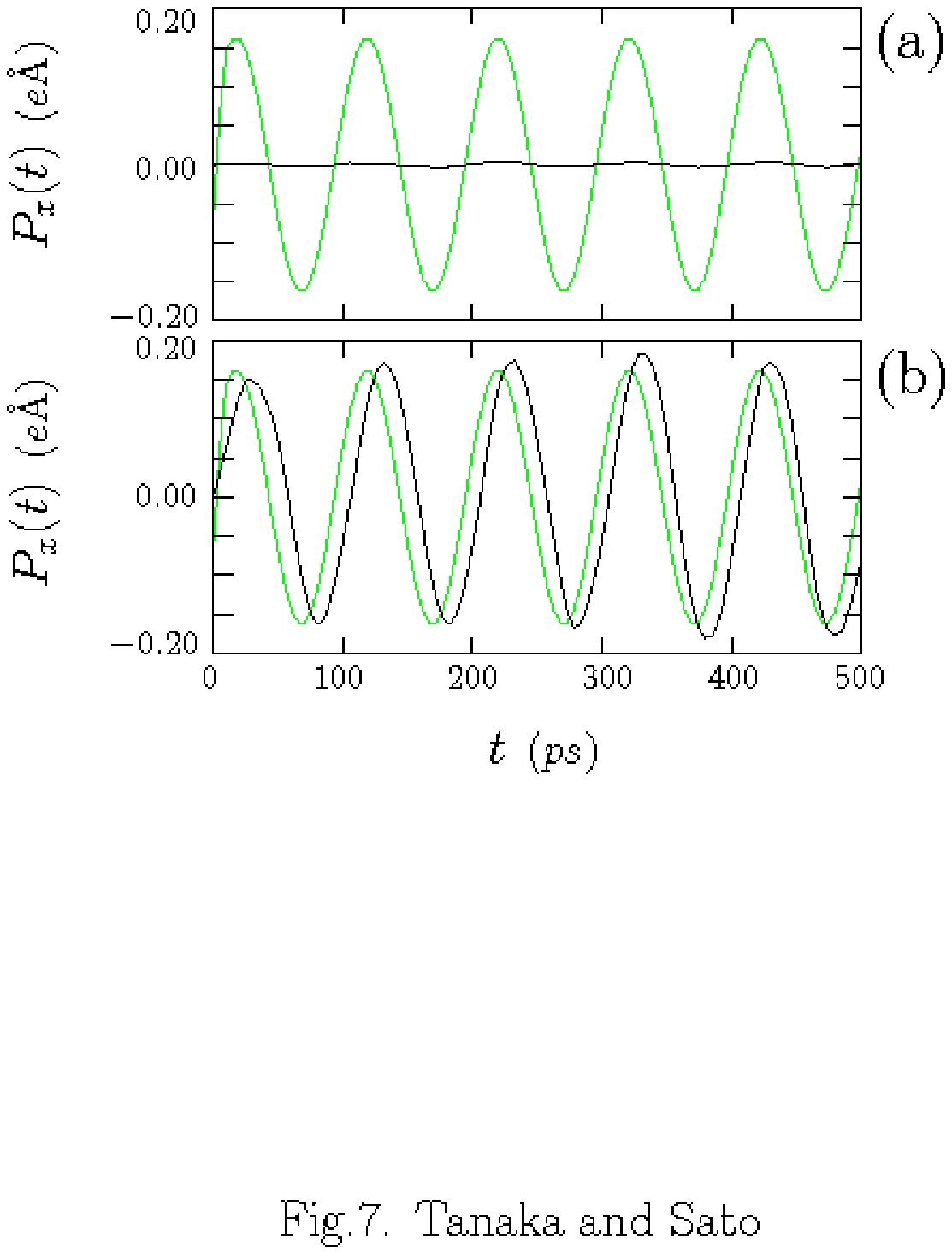}}}
\vspace*{-7.0cm}
\caption{
\noindent
The $ x $-component of the average electric dipole
$ <\hspace*{-0.1cm} P_{x}(t) \hspace*{-0.1cm}>= 
\sum_{i} {\bf P}_{i} \cdot \hat{x}/N $ and the microwave 
electric field $ E_{x}(t) $ (gray lines) for (a) the ice 
at 230K shown in Fig.1, and (b) liquid water initially at 
300K shown in Fig.4.
}
\label{Fig.7}
\end{figure}

\newpage
\begin{figure}
\vspace*{-5.0cm}
\centerline{\scalebox{0.80}{\includegraphics{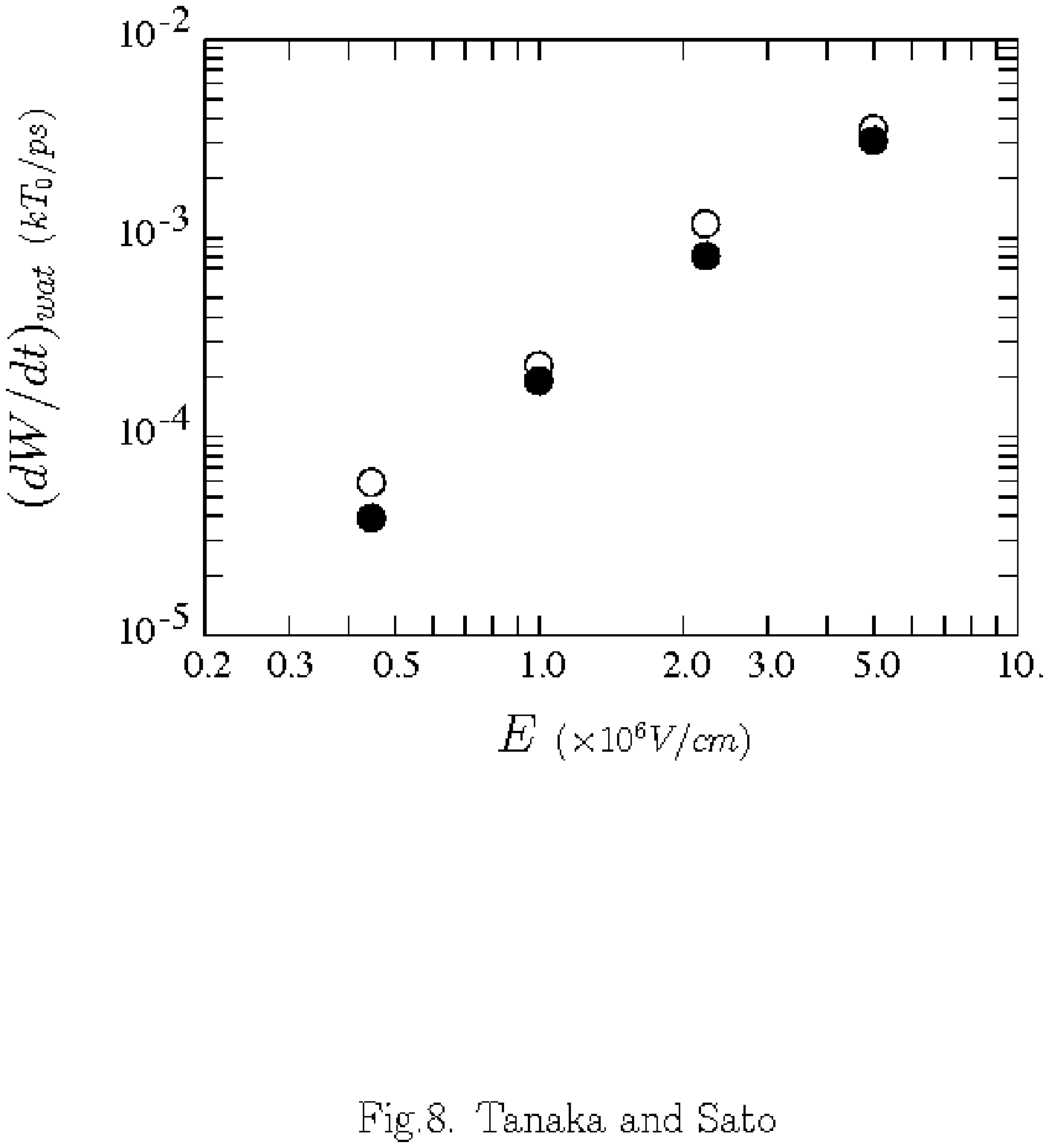}}}
\vspace*{-5.0cm}
\caption{
\noindent
The dependence of the energy transfer rate
from microwaves to liquid water (initially at 300K), on the 
strength of microwave electric field is shown for the kinetic 
energy $ dW_{kin}/dt $ (filled circles), which corresponds to 
heating of water, and the system total energy 
$ dW_{sys}/dt $ (open circles), which includes the kinetic 
and inter-molecular energies.  The microwave frequency is 10GHz. 
The data points are well fit by power laws 
Eq.(\ref{eq:dT/dt-E-sim}).
}
\label{Fig.8}
\end{figure}

\newpage
\begin{figure}
\vspace*{-5.0cm}
\centerline{\scalebox{0.80}{\includegraphics{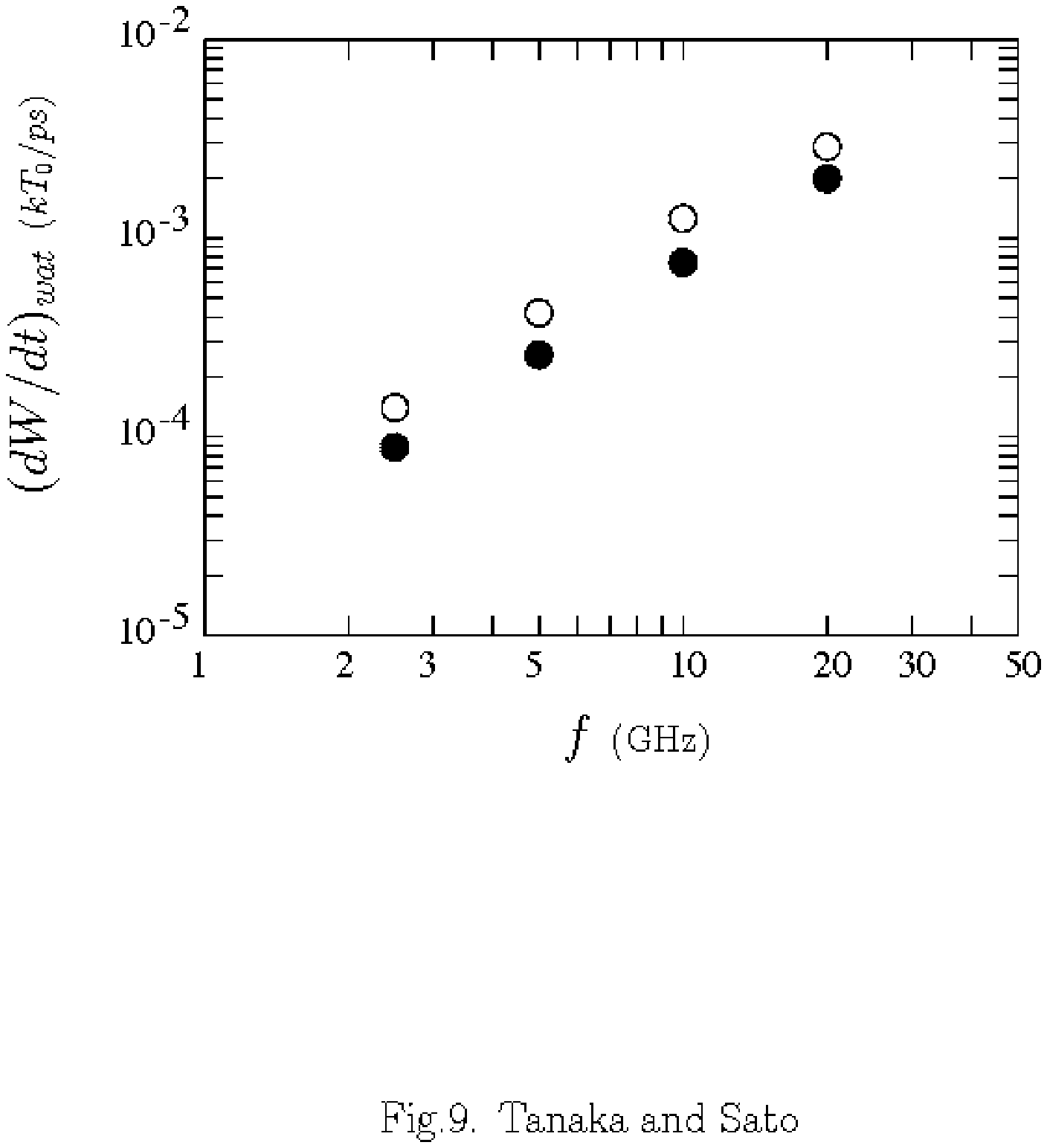}}}
\vspace*{-5.0cm}
\caption{
\noindent
The dependence of the energy transfer rates from 
microwaves to liquid water (initially at 300K), on the 
frequency $ f $ (GHz) of microwaves is shown for the 
kinetic energy $ dW_{kin}/dt $ (filled circles), and 
for the system total energy $ dW_{sys}/dt $ (open circles).
The electric field strength is $ 2.23 \times 10^{6} $V/cm
(or $ E_{0}p_{0}/kT_{0} \sim 0.42 $).  
The data points are well fit by the power laws 
Eq.(\ref{eq:dT/dt-f-sim}).
}
\label{Fig.9}
\end{figure}

\newpage
\begin{figure}
\vspace*{-3.0cm}
\centerline{\scalebox{0.80}{\includegraphics{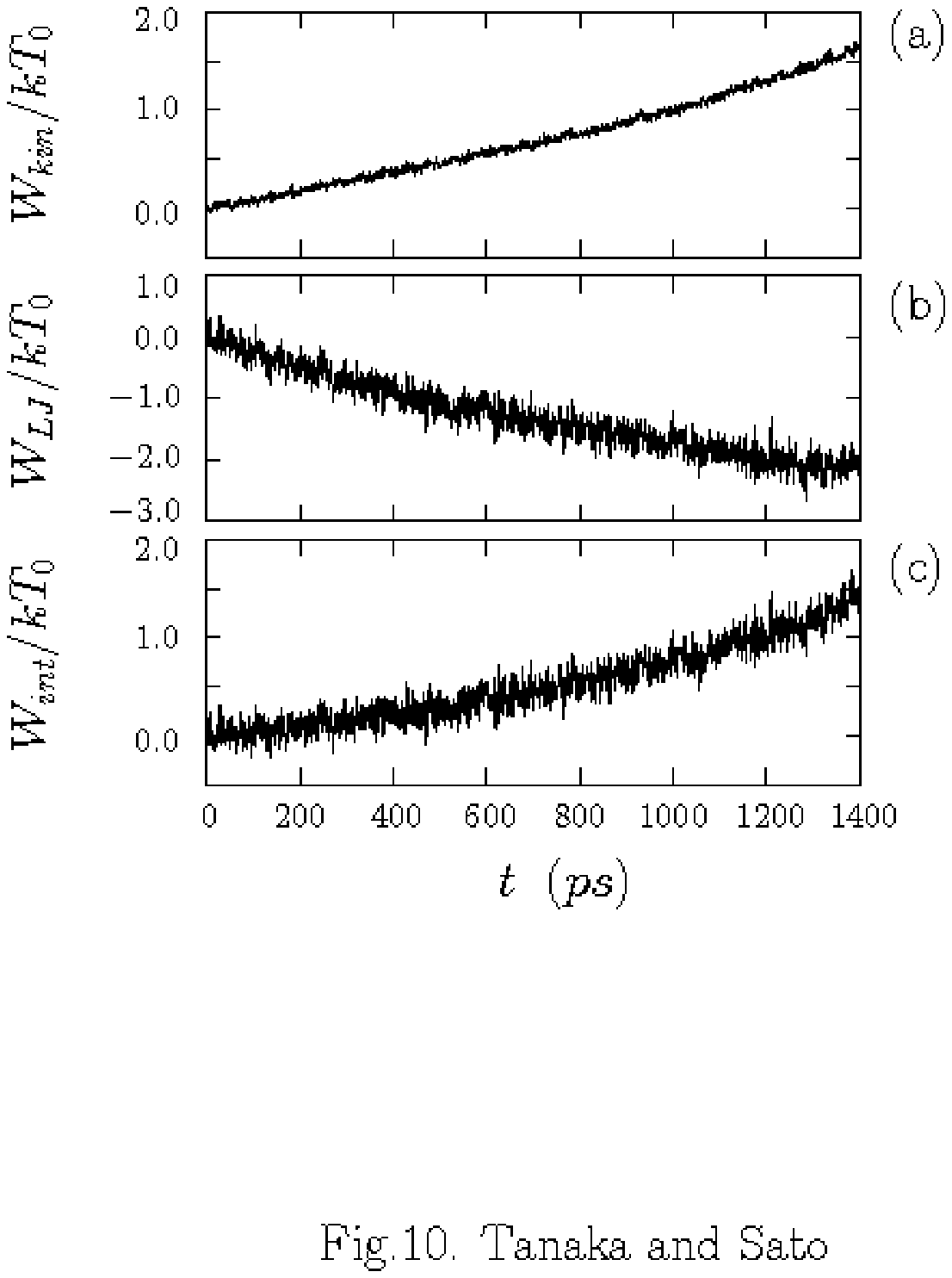}}}
\vspace*{-7.0cm}
\caption{
\noindent
The time history of (a) the average kinetic energy, (b) 
the average Lennard-Jones energy, and (c) the sum 
of average Coulombic and Lennard-Jones energies per molecule, 
for salt water with 1mol\% NaCl salinity initially at temperature 
300K.  Microwave frequency is 10GHz and its strength is 
$ E_{0}p_{0}/kT_{0} \sim 0.42 $.
}
\label{Fig.10}
\end{figure}

\newpage
\begin{figure}
\vspace*{-3.0cm}
\centerline{\scalebox{0.80}{\includegraphics{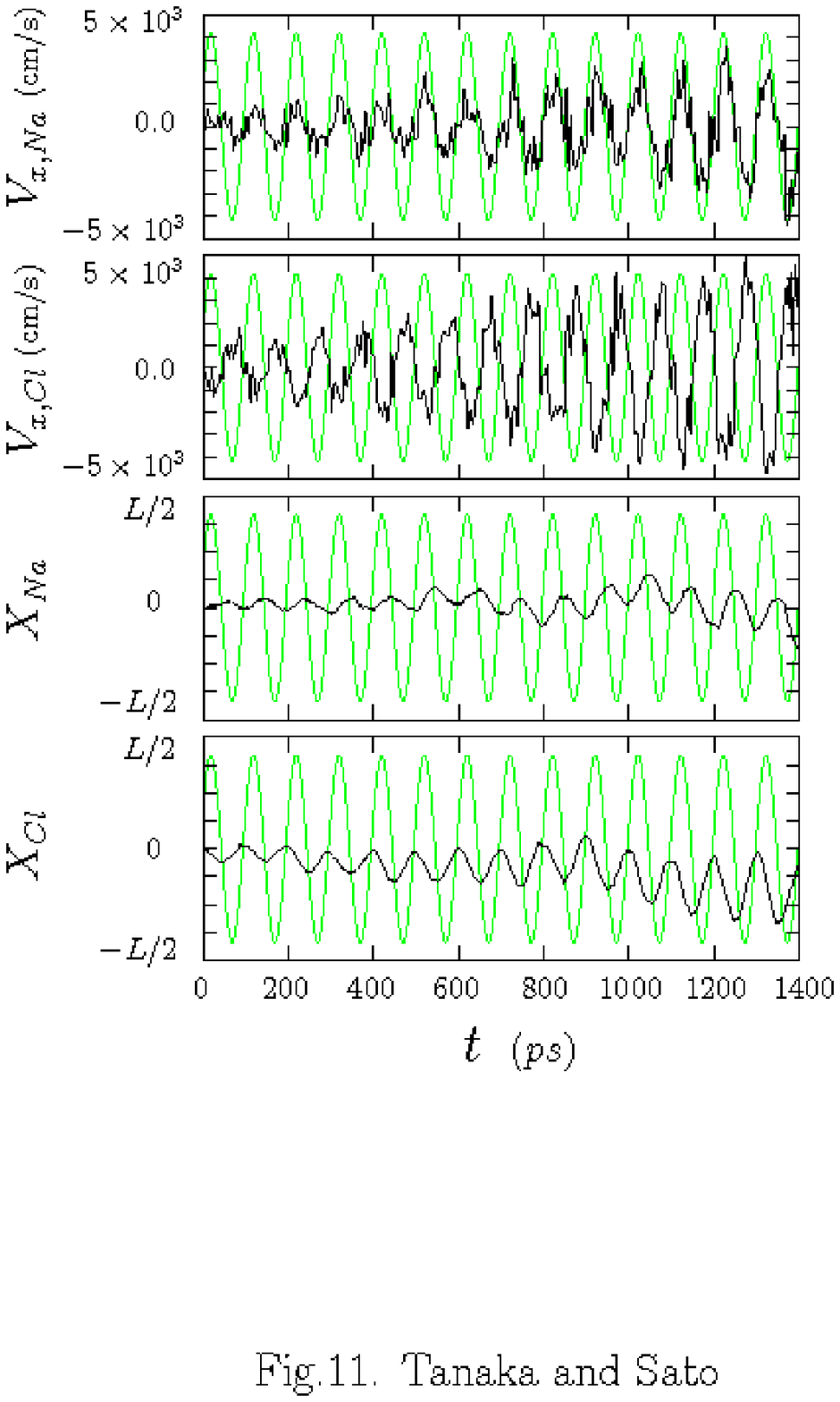}}}
\vspace*{-5.0cm}
\caption{
\noindent
The time history of the average $ x $-component velocity of 
Na$^{+}$ and Cl$^{-}$ ions, and that of the average positional shift 
of Na$^{+}$ and Cl$^{-}$ ions in the $ x $ direction (from top to 
bottom), for the run shown in Fig.10.
Gray lines are the amplitude of the applied electric field.
}
\label{Fig.11}
\end{figure}

\newpage
\begin{figure}
\vspace*{-2.0cm}
\centerline{\scalebox{0.80}{\includegraphics{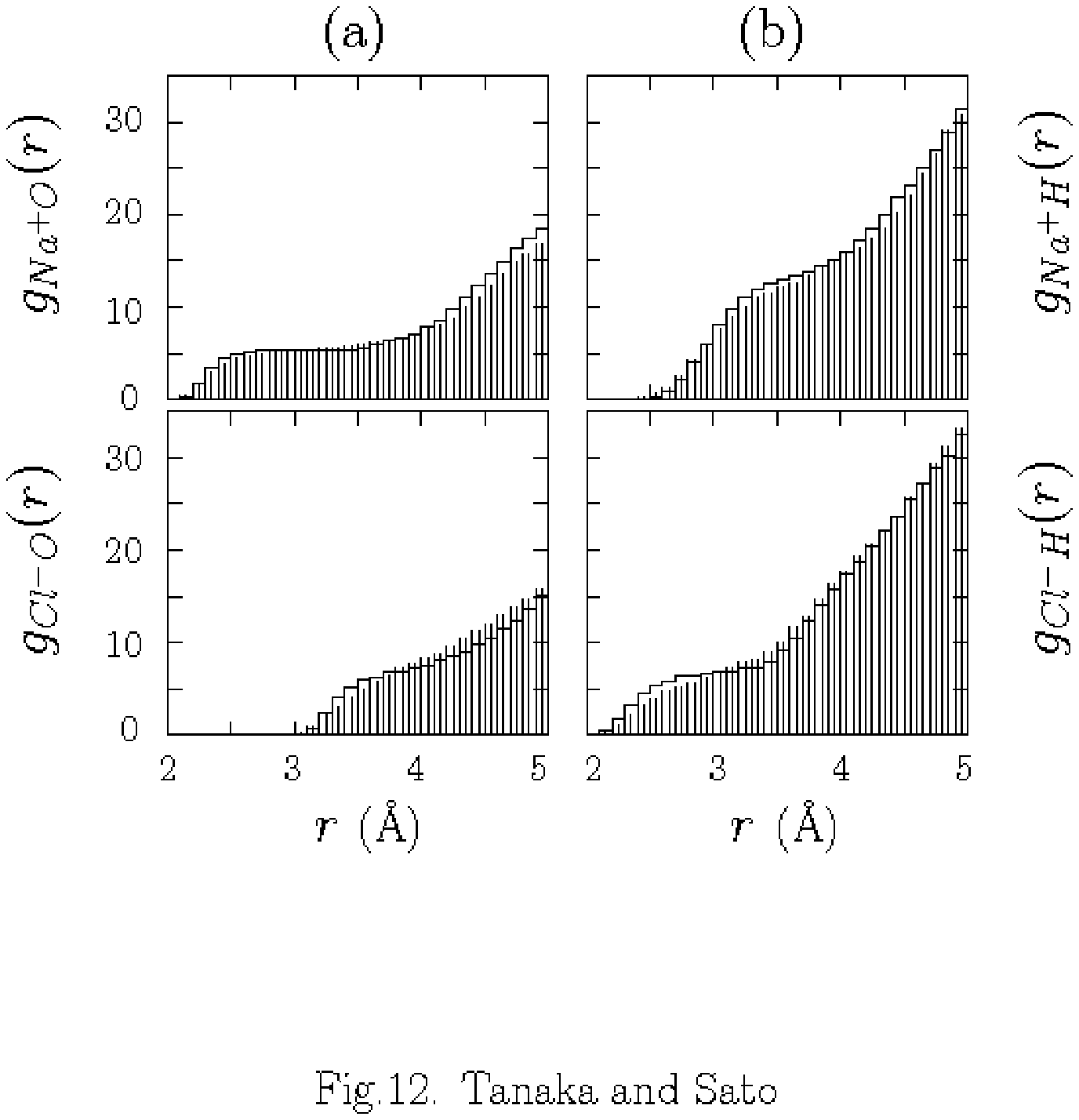}}}
\vspace*{-7.0cm}
\caption{
\noindent
The accumulated radial distribution functions of 
Na$^{+}$ (top row) and Cl$^{-}$ ions (bottom row) with respect to 
(a) oxygen atoms and (b) hydrogen atoms, for the initial time 
$ t= 0 $ (solid line) and after the microwave application at 
$ t= $ 1.4ns (shaded histogram) for the salt water shown in Fig.10.
Microwave frequency is 10GHz and its strength corresponds 
to $ E_{0}p_{0}/kT_{0} \sim 0.42 $.
}  
\label{Fig.12}
\end{figure}

\newpage
\begin{figure}
\vspace*{-5.0cm}
\centerline{\scalebox{0.80}{\includegraphics{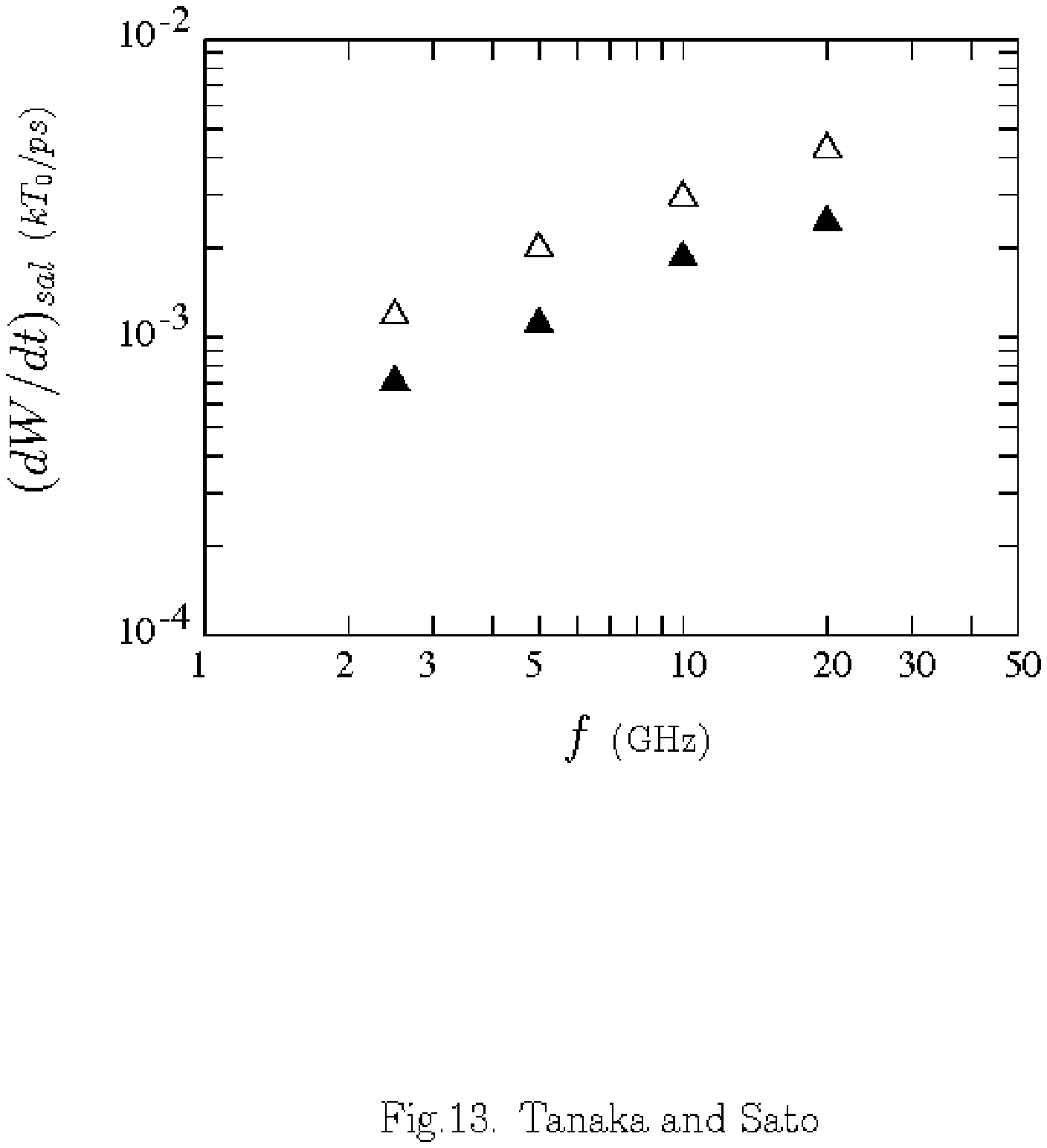}}}
\vspace*{-5.0cm}
\caption{
\noindent
The dependence of the energy transfer rate from microwaves to 
salt water with 1mol\% salinity, on the frequency $ f $ (GHz) 
of microwaves is shown for the kinetic energy $ dW_{kin}/dt $ 
(filled triangles), and for the system total energy 
$ dW_{sys}/dt $ (open triangles).
The electric field strength is $ 2.23 \times 10^{6} $V/cm, 
which corresponds to $ E_{0}p_{0}/kT_{0} \sim 0.42 $.
}
\label{Fig.13}
\end{figure}


\begin{thebibliography}{}
\bibitem{iceIc}
Initial configuration of ice is due to courtesy of Dr. M.Matsumoto.
\end{thebibliography}
\end{document}